# Anisotropic Connectivity and its Influence on Critical Current Densities, Irreversibility Fields, and Flux Creep in In-Situ-Processed MgB$_2$ Strands

Z. X. Shi [a, b, *], M. A. Susner [a], M Majoros [a], M. D. Sumption [a, **], X. Peng [a,c], M. Rindfleisch [c], M.J. Tomsic [c] and E.W. Collings [a]

[a] CSMM, Department of Materials Science and Engineering, The Ohio State University, Columbus, OH 43210, U.S.A.

[b] Department of Physics, Southeast University, Nanjing 211189, P. R. China

[c] Hyper Tech Research Inc., Columbus, OH 43212, U.S.A.

Corresponding authors: *zxshi@seu.edu.cn, **sumption.3@osu.edu

**Abstract**

The anisotropy of the critical current density ($J_c$) and its influence on measurement of irreversibility field ($B_{irr}$) has been investigated for high quality, in-situ MgB$_2$ strands. Comparison of transport and magnetization measurements has revealed the onset of a regime where large differences exist between transport and magnetically measured values of the critical current density and $B_{irr}$. These effects, initially unexpected due to the lack of crystalline texture in these in-situ processed strands, appear to be due to a fibrous microstructure, connected with the details of the wire fabrication and MgB$_2$ formation reactions.

Scanning electron micrographs of in-situ-processed MgB$_2$ monocore strands have revealed a fibrous microstructure. Grains (~100 nm) are randomly oriented, and there is no apparent local texture of the grains. However, this randomly oriented polycrystalline material has a fibrous texture at a larger length scale, with stringers of MgB$_2$ (~ 60 μm long and ~5 µm in diameter) partially separated by elongated pores -- the spaces previously occupied by stringers of elemental Mg. This leads to an interpretation of the differences observed in transport and magnetically determined critical currents, in particular a large deviation between the two at higher fields, in terms of different transverse and longitudinal connectivities within the strand. The different values of connectivity also lead to different resistive transition widths, and thus irreversibility field values, as measured by transport and magnetic techniques. Finally, these considerations are seen to influence estimated pinning potentials for the strands.

1. Introduction

Magnesium diboride (MgB$_2$) superconductors have attracted considerable interest over the years, first because of their novel two-band superconductivity, and later because of the possibility of achieving large upper critical fields ($B_{c2}$) in practical strands, based on the early thin-film results, e.g. [1][2]. Although both $B_{c2}$ and transport critical current density ($J_c$) have been significantly improved by elemental substitution, nanoparticle doping, and irradiation [3][4][5], the $J_c$ of the typical powder-in-tube strand still drops prematurely with increasing magnetic field. This is due to a combination of several factors, viz: crystalline anisotropy [6][7], a lack of a high density of strong pinning centers, a suppressed upper critical field that limits the operating scope of the material, an intrinsically low density, and insufficient connectivity between the grains [8][9][10]. Additionally, in high fields near $B_{c2}$, the magnetic relaxation rate of MgB$_2$ appears to increase rapidly, which is inconvenient for application [11].

In investigating some of these properties for a sample of in-situ-processed PIT MgB$_2$ strand, measurements were made of: (i) the magnetic $J_c$ in perpendicular and parallel fields, (ii) the perpendicular-field transport $J_c$, (iii) the temperature dependence of the resistance in perpendicular and parallel applied magnetic fields, and (iv) magnetization decay. In comparing the field dependencies of transport- and magnetic $J_c$s ($J_{ct}$ and $J_{cm}$, respectively) it was noted that although agreement was obtained at low applied field strengths, as the field increased beyond about 8 T (at 4.2 K) $J_{cm}$ began to drop much faster than $J_{ct}$. An inquiry into this discrepancy stimulated the present research which evolved into an investigation of anisotropy using comparative measurements of transport and magnetic measurements. It was seen that the strand had an underlying anisotropy in $J_c$ and that this led to consequences not only for comparisons of magnetic and transport $J_c$, but also to differences in the estimations for irreversibility fields as determined magnetically and resistively, and finally to an influence on the effective pinning potential, $U_{eff}$, estimated using different techniques.

Finally a note on terminology: In an aid to clarity, the two cardinal directions of the applied field with respect to the strand axis are designated "perpendicular" and "parallel", respectively. Those of current *within* the strand are designated "transverse" and "longitudinal".

## 2. Experimental

*2.1 Sample Preparation:* Monofilamentary powder-in-tube (PIT) strands 0.834 mm in diameter with a Nb chemical barrier and a Monel outer sheath ($MgB_2$/Nb/Monel) were manufactured by Hyper Tech Research, Inc. (HTR) using a previously described process [12][13]. The starting powders were Mg (99%, 20-25 µm maximum size), B (99.9%, amorphous, 1-2 µm maximum agglomerate size), and SiC (30 nm). With a final stoichiometry of $Mg_{1.15}B_2$+5 mol%SiC in mind appropriate masses of the Mg, B, and SiC powders were V-mixed and then ball-milled in preparation for tube filling. The strand, tracer number 1205, consisted of 25.7% $MgB_2$ core (by area), 26.2% Nb, and 48.1% Monel. Straight samples approximately 30 cm long (with crimped ends) were reacted in flowing argon according to the following schedule: ramp to 700 °C in 45 min, hold for 40 min, cool to room temperature (within approximately 45 min).

*2.2 Measurements:* Magnetic, transport, and electron optical techniques were used to examine the MgB$_2$ strand described above. First, transport and magnetic measurements were made on samples 5-15 mm long using a Quantum Design Model 6000 "physical property measuring system" (PPMS). They consisted of: (i) measurements of magnetic $J_c$ (i.e. $J_{cm}$) at 4 to 40 K extracted from the *M-H* (magnetization vs applied field) loops for samples in perpendicular and parallel fields, (ii) *R* vs *T* measurements from about 4-40 K (a) in perpendicular fields of up to 14 T using 10 Hz AC currents of 14 mA (10 A/cm$^2$) and 140 mA (100 A/cm$^2$), and (b) in perpendicular and parallel 5 T applied fields at 5 mA DC, (iii) Magnetization vs Temperature (*M-T*) curves taken from 4 to 40 K in perpendicular fields of 10 mT to 9 T, and (iv) magnetization decay (flux creep) measurements of effective pinning potential at 20 K.

In addition, using different test equipment, transport measurements of $J_c$ (i.e. $J_{ct}$) were made on strand samples 3 cm long (voltage tap separation 3 mm) in perpendicular magnetic fields of up to 14 T. The sample probe used a He exchange gas and conduction cooled current leads in conjunction with a heater and temperature controller to set the measurement temperature. The probe could pass measuring currents of up to 150 A with negligible sample heating. Results were obtained for a range of critical current (electric field) criteria ranging from 1 µV/cm to as low as 0.0005 µV/cm.

Finally, longitudinal and transverse microstructures of the strand were examined using an FEI-Sirion ultrahigh resolution scanning electron microscope (SEM), and used to interpret the magnetic and transport results.

## 3. Results

### 3.1 *Scanning Electron Microscopy*

Figure 1 compares the (a) longitudinal and (b) transverse sections of the reacted core of the present in-situ-processed monocore strand and reveals a pronounced anisotropic (fibrous) microstructure consisting of stringers up to 60 µm long and about 5 µm thick separated by elongated pores. Figure 1(c) shows the stringers to consist of randomly oriented polycrystalline $MgB_2$ grains ~ 100 nm in size. The formation of a fibrous as-drawn macrostructure followed by an elongated porous structure after reaction heat treatment can be understood with reference to Giunchi's reports on: (i) the processing of long $MgB_2$ tubular wires by drawing followed by the "reactive liquid infiltration" (RLI) of Mg [14], and (ii) the RLI processing of bulk $MgB_2$ shapes such rods, tubes, etc [15].

Pure Mg, with an elongation-to-fracture of ~7% under static strain, as compared to those of Cu (~54%) or even Fe (~17%) [16] is well known for its lack of ductility. It was then indeed surprising to discover that when imbedded in fine B powder (either amorphous or microcrystalline) an axial rod of Mg could be co-reduced by drawing to large area reductions. Both Giunchi [14] and later Kumakura et al. [17] have succeeded in elongating to wire small experimental billets of Mg rod packed in B. For this reason the elongation to stringers of the Mg-particle component of a pre-reacted Mg-plus-B PIT wire core can be expected to take place under similar circumstances. The replacement of the Mg stringers with elongated vacancies can be understood with reference to the results of Giunchi's experiments on the RLI processing of $MgB_2$ bulks. In these it was found that the otherwise fully dense $MgB_2$ formed during the heat treatment of a piece of Mg imbedded in compacted B powder contained a vacant space whose shape exactly replicated that of the original Mg [15].

Based only on the fact that the grain orientations are random (there is no grain orientation based texture), we would expect a strand with relatively isotropic effective electromagnetic properties. Thus, while $MgB_2$ is intrinsically anisotropic, the lack of preferential grain orientation should lead to an apparently isotropic strand at length scales significantly larger than the grain size (~100 nm). On the other hand, there is a texture within the strand at a larger length scale (see Figure 1 (a)). This microstructural texture as shown in Figure 1 (a) (longitudinal) and (b) (cross section) is one based on porosity and connectivity. The strand thus can be considered to be constructed from "fibers" which are themselves assembled from relatively well connected grains which are nonetheless random in orientation. The fibers have a preferential longitudinal alignment within the strand, and are connected to one another, but with the possibility for fiber-to-fiber connectivity levels which are different transversely and longitudinally within the strand. Differences between magnetic and transport $J_c$, as well as differences in transverse and longitudinal critical currents, will be shown below. These differences are attributed primarily to a "porosity" or "connectivity" texture, rather than one based on grain texture. This will influence measurements of critical currents, irreversibility fields, and pinning potentials.

### 3.2 *Magnetic and Transport Critical Current Densities*

Measurements of magnetization were made on cylindrical strand samples at several temperatures from 4.2 K to 40 K in fields of up to 14 T applied perpendicular and parallel, respectively, to the strand axis at a sweep rate of 13 mT/s. Values of perpendicular and parallel $J_{cm}$s were determined from full hysteresis-loop heights, $\Delta M$, using the standard Bean-model-based expressions (in SI units) where $J_{cm}$(perpendicular) = $(3\pi/8)\Delta M/R$ and $J_{cm}$(parallel) = $(15/10)\Delta M/R$ in which $R$ is the (mono)filament radius.

Figure 2 displays $J_{cm}$ vs $B$ at various temperatures for fields both parallel and perpendicular to the strand. $J_{cm}$ as measured in perpendicular applied fields, $J_{cm\perp}$ is approximately two times larger than $J_{cm}$ as measured in parallel applied fields, $J_{cm//}$, except at higher fields, where the $J_c$ curve is not exponential (not linear on a log-linear plot). Beyond some point as the field increases (a point which decreases with increasing temperature), the magnetization drops rapidly for both orientations, meeting at a point which would appear to correspond to an irreversibility field.

If the anisotropic macrostructure shown in Figure 1 is indeed accompanied by anisotropic connectivity this would explain the observed field-dependent differences between the perpendicular and parallel $J_{cm}$s. Indeed, Figure 1 shows these stringers to be porous and separated by a labyrinth of vacancies. It is plausible to suggest such macroscopic structural anisotropy leads to anisotropic connectivities, associated with the longitudinal and transverse directions in the strand, respectively. A cartoon, Figure 3, illustrates the results (see also Ref [18,19,20]). Let us start by assuming a transverse critical current, $J_{cT}$, which is less than the longitudinal value $J_{cL}$. Let us also assume that $J_{cT}$ drops off more rapidly in field. In this case, the relevant magnetization expressions are given by the case of the anisotropic cylinder [18]. In general, the magnetization is controlled by the smaller of two quantities; $J_{cL}*d$ or $J_{cT}*L$, where $d$ is the diameter of the superconductor and $L$ is the length. The two limiting cases are simple: (i) At low fields, $J_{cT}$ is at its largest relative to $J_{cL}$, and while $J_{cT} < J_{cL}$, it is also true that $L >> d$, such that $J_{cT}*L > J_{cL}*d$, allowing sufficient length for the current to transfer across the strand at the ends (Kirchhoff's law requires the current loops to close), and thus the field-perpendicular magnetization is proportional to $dJ_{cL}$. This situation is represented at left in Figure 3. (ii) At high fields, $J_{cT}$ has dropped considerably, and flux penetration along the sample length, due to the now weak $J_{cT}$, causes the field-perpendicular magnetization to be proportional to $J_{cT}$, as seen in the middle of Figure 3. Of course, in parallel field, the $J_c$ is always controlled by the $J_{cT}$ (as shown at right in Figure 3) which we assume to be uniform in the radial and tangential directions. This is consistent with the results seen in Figure 2 where the low field magnetic $J_c$ for perpendicular fields is higher than that for parallel fields, but as the field is increased, the two magnetic $J_c$s coalesce and finally drop together rapidly at higher fields.

We can formalize our model for explaining these effects based on the following elements: (i) an "observed" (i.e., effective) longitudinal critical current density $J_{cL}(B,T) = K_L(B,T) \cdot j_c(B,T)$ where $K_L$ is the longitudinal connectivity and $j_c$ is the average intra-stringer critical current density (taken to be averaged over lengths scales significantly larger than the grain sizes and thus effectively isotropic); (ii) an "observed" transverse critical current density, $J_{cT}(B,T) = K_T(B,T) \cdot j_c(B,T)$ where $K_T$ is the transverse connectivity; (iii) the assumption that a relatively weak stringer-to-stringer interconnection exists, Figure 1(a), such that $K_T(B,T)$ is both smaller than and more strongly field-dependent than $K_L(B,T)$.

Figure 4 shows an interesting comparison of magnetic and transport based $J_c$s, and there is a clear tendency for them to agree quite well as low fields, and then at some field the magnetization derived value falls away from the transport value rapidly. Note that the temperature ranges from 4-25 K for the data of Figure 4, but the bifurcation of the magnetic and transport $J_c$s at higher fields are seen at all temperatures (except 25 K, see below).

In comparing magnetic and transport derived $J_c$s, it is important to note that the value of $J_c$ is affected by the selected electric field criterion due to flux creep. For transport measurements, the difference in values of $J_c$ resulting from varying the electric field criterion is large. Correspondingly, for magnetic measurement, *M-H* loop heights (and the derived $J_c$s) are influenced by the applied field sweep rate. For *M-H* loop measurements in the present work, the field sweep rate was 13 mT/s, corresponding to a voltage criterion of 0.04 µV/cm for $J_{cm}$, a value which is about 25 times less than the $J_{ct}$ criterion of 1 µV/cm. In order to asses the effect of the difference in voltage criterion for $J_{cm}$ and $J_{ct}$, we re-calculated $J_{ct}$ using various voltage criteria, ranging from 1 µV/cm down to 0.0005 µV/cm, and the results are displayed in Figure 5. Here we can see that there are substantial changes in $J_c$ (as well as the apparent $B_{irr}$) as expected – but nowhere near enough to explain the large and systematic differences between $J_{cm}$ and $J_{ct}$ at high fields as seen in Figures 4 and 5. In point of fact, $J_{ct}$ determined by a criterion of 0.0005 µV/cm is still much higher than $J_{cm}$ at high fields. The fact that the electric field corresponding to $J_{cm}$ (estimated to be ~ 0.04 µV/cm) lies well among the values quoted above confirms that the high field divergence of $J_{cm}(B)$ and $J_{ct}(B)$ is not measurement-criterion related.

Figure 6 shows an illustration similar to Figure 3, in this case intended to illustrate the differences between the measured transport- and field-perpendicular magnetic $J_c$s. As stated above, at low fields the field-perpendicular magnetization is controlled by $J_{cL} = K_L \cdot j_c$ such that the $J_{ct}(B)$ and $J_{cm}(B$-perpendicular$)$ curves coincide. At high fields the field-perpendicular magnetization is dominated by $J_{cT} = K_T \cdot j_c$ and hence yields a value $J_{cT}(B)$ which therefore falls below the transport measured $J_{ct}(B)$.

Our simple model thus has several virtues. Based on the simple assumption of an anisotropic connectivity, strongly suggested in fact by SEM micrographs, a model emerges which explains both the difference in parallel and perpendicular magnetic $J_c$s and their field dependence relative to one another, but also explains a curious falling off of the measured magnetic $J_{cT}$ from the transport $J_{ct}$. The model seems to describe the results of Figures 2, 4, and 5 relatively well, although at 25 K in Figure 4 we see that the magnetic $J_c$ exceeds the transport, this implies some more complicated temperature dependence of the connectivity (which we will avoid attempting to describe at this point). However, if this connectivity model is correct, it must have implications for other aspects of the strand's electromagnetic properties. In fact, on the face of it, Figure 5 would suggest a very different value for the irreversibility field depending upon whether it was measured with transport or magnetic methods. Below we investigate this further.

### *3.3 Transport Measurements of the Critical Fields and Irreversibility Fields*

Resistance vs temperature curves were measured from 4 to 40 K in perpendicular and parallel applied fields using a DC current of 5 mA as well as 10-Hz AC currents of 14 mA and 140 mA. Measurements were repeated for a range of field strengths of up to 14 T. A typical set of results is shown in Figure 7, in this case measured at 5 T. Figure 8 displays the upper and lower transition points of the $R(T)$ curves. Below we have denoted the resistive measurements in generally as $B_{c2}{}^r$ and $B_{irr}{}^r$, but in Figure 8 we have used $B_{c2}{}^r{}_{(AC-RES)}$ and $B_{irr}{}^r{}_{(AC-RES)}$ to further specify that the measurements were made using an AC current profile. AC and DC measurements of critical fields gave essentially identical results, irrespective of field, although different AC current levels (specifically 14 mA vs 140 mA) did change the $B_{irr}$ values somewhat. Several observations can be made:

(a) *Upper Transition:* $B_{c2}$ values determined from either the DC or AC $R(T)$ curves (as the point at which the resistance begins to drop with decreasing temperature -- arbitrarily taken as 90% of the normal value) give a similar $B_{c2}$ value, as expected. For this specific set of $R$-$T$ curves, $B_{c2} = 5$ T at 26.2 K. The full set of $B_{c2}$ values is displayed in Figure 8. The $B_{c2}$ in question is presumably similar to $B_{c2}{}^{ab}$, since when the first grains become superconducting (dropping from a higher field or temperature), there will be a drop in the total resistance, even while some grains along the path are not in the superconducting state. However, the magnetic results give slightly higher values, as will be commented on further below.

(b) *Lower Transition:* The position of the occurrence of the lower transition (here taken as 5% of the normal state resistivity) and the resulting finite width of the resistive transition has been attributed to various effects or combinations of them. A key contribution is of course the intrinsic anisotropy in $B_{c2}$ which as shown by Eisterer [6] provides a contribution to the $R(T)$-transition width, $\Delta T_a$, given by

$$\Delta T_a = \frac{\left(\sqrt{(\gamma^2 - 1)p_c^2 + 1} - 1\right)}{-\partial B_{c2}/\partial T} B_0 \tag{1}$$

in which $\gamma$ is the upper critical field anisotropy factor, $B_{c2}{}^{ab}/B_{c2}{}^c$ (in which $B_{c2}{}^{ab}$, or $B_\perp$, is the critical field for *B* parallel to the *ab*-plane and $B_{c2}^{c}$ is that for *B* parallel to the *c*-axis) $p_c$ is the percolation threshold, $B_0$ denotes the external field, and it is assumed that $\gamma$ and $\partial B_{c2}/\partial T$ are constant during the transition. Beyond this, however, there is a contribution to $\Delta T$ due to the difference between $B_{c2}$ and the irreversibility field, $B_{irr}$ [4][21]. In fact, in a condition where the field is applied transversely to the current, the lower value ***must*** be associated with $B_{irr}$, since otherwise significant flux flow resistivity would be present. For $MgB_2$ strands there is usually not a strong difference between the transition curves in configurations where the field is applied parallel or perpendicular to the current, presumably because the percolative path of current flow causes there to be both parallel and perpendicular current/field orientations no matter what the nominal current and field directions are.

Thus, the lower transition is associated with $B_{irr}$ and the upper transition $B_{c2}^{ab}$, even though a significant portion of the transition width is due to anisotropy in $B_{c2}$. Also contributing, at some level, is thermally activated flux flow (TAFF) [22] which is particularly pronounced for grains whose c-axes are oriented parallel to the applied field direction [23][24]. There are of course other possible contributions, such as sample inhomogeneity or internal strain that in comparable bulk samples has manifested itself as a broadening of the electronic-specific-heat transition [9].

(c) ***Field Orientation Independence of Transition:*** The independence of *R(T)* to the direction of the applied field indicates that the superconductor, although present in the form of highly aspected stringers, is effectively isotropic on length scales larger than the grain size but smaller than the stringer diameter (the micron length scale)-- as expected from the fine randomly oriented granular substructure depicted in Figure 1(c). While this independence of orientation might suggest that the contribution of TAFF is small, since this is an effect associated with flux flow, it is important to again account for the percolative nature of the current flow. Under this condition, both configurations where the current is perpendicular and parallel to the field will be present, as noted above.

(d) *Current Excitation Dependence of the Lower Transition (and thus transition width):* According to Figure 7 the total width of the resistive transition is 1.1-1.6 K depending on the measuring current, 5-140 mA, respectively. The $B_{c2}$ anisotropy contribution must of course be less than this, which restricts the allowable ranges of $p_c$ and $\gamma$. We can first consider the percolation threshold. As the temperature, $T$, of a system of randomly oriented grains is lowered in the presence of a "sensing field", say $B_0$ applied in any direction, a percolation path eventually develops (through what is now a system of superconducting and normal particles) as $B_{c2}^c(T)$ emerges above $B_0$. So-called "bond" and "site" percolation through resistive networks and conducting/insulating composites has been the subject of extensive study and review [25][26] as a result of which $p_c$s ranging from 0.15 [27][28] to 0.3 ± 0.1 [25][29] have been obtained. The latter group of values is based on the results of overlapping-sphere- and simple-cubic resistor- network calculations. Accordingly, for inclusion in our estimate of $\Delta T_a$ we select $p_c = 0.3$, which in fact is a number previously used in calculations of transport-current percolation through porous $MgB_2$ [8]. Next, we can consider the anisotropy ratio, $\gamma$ which has also been extensively studied. Based on measurements of films, single crystals, aligned crystals, and other samples, $\gamma$-values ranging from 1.1-1.7 [30][31][32] through 1.8-2.0 [33] to 3-7 [23][34] have been deduced. If anisotropy is responsible for the entire width of the resistive transition (1.6 K), Equation (1) yields $\gamma = 2.7$. However, the lower transition must be associated with $B_{irr}$, and thus the transition width due to anisotropy must be less than this total amount. To the extent that some level of TAFF is also present, this requires a somewhat smaller $\gamma$ still.

With the above considerations we have taken some care to examine the transport based estimations of $B_{irr}$ and $B_{c2}$. This alone is not particularly new data, but of interest is the direct comparison of this data (taken via transport measurement) to measurements of the same quantities by magnetic technique.

### 3.4 *Magnetic Measurements of Critical Fields and Irreversibility Fields*

We note that there are several approaches to $B_{irr}$ measurement: (a) the above-mentioned resistive *R(B)* or *R(T)* method in which $B_{irr}(T)$ is derived from some lower point of the resistive transition [35]; (b) a critical-current method in which $B_{irr}$ is taken as the fields at which $J_{ct}(B)$ or $J_{cm}(B)$ drop to some arbitrarily small values (e.g. 100 A/cm$^2$ [36] or 10 A/cm$^2$ [37], respectively); (c) the Kramer method in which $B_{irr}$ is obtained from the linear extrapolation to zero of $[\Delta M(\text{or } J_c)]^{1/2}B^{1/4}$ versus $B$; (d) a magnetization method in which $B_{irr}(T)$ are the fields at which the shielding- and trapping branches of a set of isothermal *M-H* loops either draw together at some small value of $J_{cm}$ (e.g. 100 A/cm$^2$ [38]) or visually coalesce within the resolution of the magnetometer [21]. An alternative magnetization method, the one adopted in this paper, holds the field constant and varies the sample temperature. Again the $B_{irr}(T)$s are taken as the coalescence points of shielding and trapping curves obtained in the following way: The sample's magnetization is measured (i) during warm-up in various fixed perpendicular fields (of 10 mT to 9 T) after it has been cooled to the starting temperature in the absence of field (i.e.

after zero-field cooling, ZFC) and (ii) after it has been cooled to starting temperature in the measuring field (after field cooling, FC).

In the previous section of the paper we used resistively determined irreversibility fields, $T_{irr}^r$ or $B_{irr}^r$, and we will now compare them to *M-T* curve based determinations, which we will denote $T_{irr}^m$ and $B_{irr}^m$. Figure 9 shows clearly the points of closure at $B_{irr}^m(T,H)$ of the ZFC (lower, shielding) FC (upper, trapping) *M(T)* curves. Two other features are also present in Figure 9: (i) A positive background contribution to the moment arising from the monel sheath of the strand, (ii) a disappearance of superconducting moment with increasing temperature beyond the bifurcation point corresponding to $T_c(B)$, or equivalently $B_{c2}(T)$. The combined results of the resistive and magnetic critical field measurements are presented in Figure 8.

The magnetically determined $B_{c2}$ is higher that that of the resistively determined $B_{c2}$. This stems from the width of the upper transition. For the magnetic measurements we have used the initial onset, with no 10% criterion, while with the transport based measurements, we have used a 10% criterion, resulting in a somewhat lower $B_{c2}^r$.

### 3.5 *Transport and Magnetic Measurements of Effective Pinning Potential, $U_{eff}$*

The $J_{ct}(B)$ data of Figure 5 were recast in the format *E* (V/cm) versus *J* (A/cm$^2$), and are plotted logarithmically in Figure 10(a). From the slopes of these lines, which clearly follow the expected relationship $E/E_0 = J/J_0^n$ (e.g. [39]), *n*-values of from 2 to 7 were derived corresponding to fields ranging from 10 to 5 T. Next, again following [39], the relationship $U_{eff} = nk_BT$, where $k_B$ is Boltzmann's constant, was invoked to obtain the $U_{eff}$ values listed in Table 1. The values so obtained are in agreement with those typically obtained on multifilamentary PIT-processed MgB$_2$ strands.

For comparison, magnetization measurements as functions of time (*t*) out to 166 min were made in fixed perpendicular fields of 1 to 5 T after ZFC to 20 K. The results are shown in Figure 10 (b). From the initial slope of the *ln(t)* dependence of the normalized shielding magnetization, $S = \partial(M/M_0)/\partial(ln(t))$, an effective pinning potential for flux creep at each of the measuring fields was calculated using the standard expression $U_{eff} = k_B T/S$ [40][41]. The values of $U_{eff}$ so obtained are also listed in Table 1.

In comparing the magnetic and transport based measurements, we must account for well known differences between $U_{eff}$ values derived these two sources. This includes some dependence on the electric field of measurement, and we note that the electric field will certainly be higher for the transport measurements than the decay measurements. A much stronger effect, however, is the dependence of $U_{eff}$ on the presence of macroscopic currents with the sample, which tends to significantly reduce $U_{eff}$ from its intrinsic value. This leads to the well know temperature dependence of $U_{eff}$ [42,43]. The $U_{eff}$ values seen here are relatively similar to that seen in other work, although the transport-measured $U_{eff}$s seem to be somewhat small that what might be expected from an extrapolation of the magnetic results, suggesting that perhaps the *n*-values from which they were derived are too small. No doubt partly contributing to the low *n*-values are the irregular cross sections of the MgB$_2$ fibers depicted in Figure 1(a), a kind of "sausaging" [44] which presumably persists into the multifilamentary arrangement.

## 4. Discussion

Magnetization- and transport measurements in perpendicular and parallel applied fields were made on a sample of in-situ type monocore MgB$_2$ PIT strand. Of particular interest in the present study of monocore PIT strand were observed differences in the field dependences of: (i) their magnetic $J_c$s as measured in perpendicular and parallel applied fields ($J_{cm\perp}$ and $J_{cm//}$, respectively, Figure 2) and (ii) their magnetic- and transport $J_c$s in perpendicular applied fields ($J_{cm\perp}$ and $J_{ct\perp}$, respectively, Figure 4). From the SEM micrographs we noted: (i) that the strand core consisted of polycrystalline MgB$_2$ grains, ~100 nm in size (Figure 1(c)) assembled in the form of long stringers up to 60 µm in length and ~5 µm thick, and (ii) that the stringers were separated by longitudinal pores or vacancies (Figure 1(b)) although interconnected by bridges.

We proposed a self-consistent interpretation of Figures 2,4, and 5 assuming: (i) a universal intra-stringer critical current density, $j_c(B,T)$, (ii) a longitudinal connectivity $K_L(B,T)$ and hence a longitudinal critical current density $J_{cL}(B,T) = K_L(B,T) \cdot j_c(B,T)$, (iii) likewise a transverse critical current density $J_{cT}(B,T) = K_T(B,T) \cdot j_c(B,T)$, (iv) a transverse connectivity, $K_T$, via the interconnects that is smaller and more strongly field dependent than $K_L$. It follows from these observations and assumptions that: (1) At *low fields*: (i) $J_{cm\perp}$ ($\propto K_L$) is greater than $J_{cm//}$ (which depends entirely on $K_T$), Figure 4, (ii) $J_{cm\perp}$ is identical to $J_{ct\perp}$, both of them being proportional to $K_L$, Figure 3. (2) At *high fields*: (i) $J_{cm\perp}$ tends to be the same as $J_{cm//}$, both of them being proportional to $K_T$, Figures 2 and 3, (ii) $J_{cm\perp}$ which weakens in response to a decreasing $K_T$ falls below the $K_L$-proportional $J_{ct\perp}$, Figures 4 and 5. Simply stated, in the transverse-field magnetic measurements a comparatively weak and more strongly field dependent transverse connectivity, $K_T$, is responsible for an "end effect" in the small aspect-ratio samples being measured. The transverse-field anomalies should be absent in very long samples. The practical implications of this result are that magnetic measurements of $J_c$ will underestimate transport values of $J_c$ at higher field.

The anisotropy in transverse and longitudinal currents also manifests itself in causing a difference between $B_{irr}^r$ and $B_{irr}^m$. We note that in order to have a substantial magnetization, currents must flow along the length of the sample, and they must also cross over at the ends, thus when the transverse currents become weak, the resulting magnetization is low. Indeed, when $J_{cT}*L << J_{cL}*d$, the magnetically determined $J_{cm} \to 0$. Thus, it is clear that there can be no difference between the upper and lower branches of an *M-H* or *M-T* curve at this point, and the irreversibility field has been reached. The practical result of this is that *M-T* measurements of $B_{irr}$ will always underestimate the point at which superconductive transport current can pass through the sample without dissipation.

The last set of measurements in this work concerned the pinning strength. Here, we have worked with $U_{eff}$, as it is straightforward to obtain, and an important point can be made using this simple quantity, even before moving to the intrinsic pinning potential. We used two methods, the first electrical, based on a measurement of the *I-V* curve, extraction of the *n*-value, and the use of $U_{eff} = nk_BT$ to obtain $U_{eff}$. This is a widely used technique, and numerous inferences have been based upon it [45]. The second measurement technique, also widely used, was based on magnetization decay [45,46]. We note that our measured values are representative of those reported in the literature [47] (in this case with an AC susceptibility technique). However, the real meaning of these results is obscured again by the level of connectivity within the strand. It should be noted that even in the "good" direction, the connectivity of in-situ $MgB_2$ strands, even though they produce some of the highest $J_c$s of MgB$_2$ strand, are typically 10% [48]. Thus, $n$-values are likely to be influenced not only by filament dimensional variation "sausaging", but also by the connectivity within the strand as a function of electric field. On the other hand, $U_{eff}$ values extracted with the use of magnetization measurements are self-evidently incorrect in the regime where there is a deviation of $J_{cm}$ from $J_{ct}$, since there is a clear prediction of a regime with no pinning where again dissipationless current flows in a transport measurement.

## 5. Conclusions

The anisotropy of the critical current density ($J_c$) and irreversibility field ($B_{irr}$) has been investigated for high quality, in-situ MgB$_2$ strands. Comparison of transport and magnetization measurements has revealed regimes where large differences exist between transport and magnetically measured values of the critical current density and $B_{irr}$. These effects appear to be due to a fibrous microstructure, connected with the details of the wire fabrication and MgB$_2$ formation reactions. This microstructure has been investigated via SEM. Grains (~100 nm) are randomly oriented, and there is no apparent local texture of the grains. However, this randomly oriented polycrystalline material has a fibrous texture at a larger length scale, with stringers of MgB$_2$ (~ 60 long and 5 µm in diameter) partially separated by elongated pores.

Several very unusual aspects of the electromagnetic response of the in-situ PIT wires were examined. First, there is a significant difference in transport and magnetically measured $J_c$ at high fields, with magnetic $J_c$ dropping to zero in regimes where transport $J_c$ is still substantially present. Secondly, there is a distinct difference between $B_{irr}$ as measured by magnetic and electrical techniques. These two effects are well described by a model with postulates anisotropic $J_c$s arising from anisotropic connectivities. This model, suggested by the appearance of the microstructure, is simply described with a minimum of parameters, and leads inevitably to the electromagnetic signatures seen. The basis of the difference is that magnetization response requires currents which travel along the transverse direction of the strands as well as longitudinally within them. Weaker and more field dependence transverse $J_c$s then lead to magnetic $J_c$ which vanish with increasing field more rapidly than transport $J_c$s which rely only on longitudinal currents. A similar effect is seen with magnetization based $B_{irr}$ measurements, since the criterion for their occurrence is that macroscopic magnetization currents are extinguished.

This leads to several practical results. First, *M-H* measurements will underestimate transport $J_c$ at high fields, while agreeing with them at low fields. Second, the irreversibility fields will be underestimated by magnetization-based measurements. Finally, $U_{eff}$ estimations were also made for the strands, based both on magnetization measurements and transport (*n*-value) measurements. It can be assumed that these values also represent an underestimation of the pinning potential. This underestimation stems not only from the usual sources of field gradients and electric field based variations, but from the following facts. In transport measurements, *n*-values are almost certainly suppressed from a connectivity reduction, leading to significantly suppressed $U_{eff}$ values (connectivity is estimated to be 10% or so in these strands). On the other hand, magnetic measurements, the magnetic $J_c$s, the decay of which we use to estimate $U_{eff}$, are actually vanishing prematurely, and do not give a good estimate of $U_{eff}$ either.

**Acknowledgements**

This work was supported by a project from the Jiangsu Province Government and by the Ministry of Education of the People's Republic of China, NCET-05-0461, and (for OSU) by a grant from the DOE office of High Energy Physics, Grant No. DE-FG02-95ER40900.

anisotropy from magnetization data on random powders as applied to LuNi$_2$B$_2$C, YNi$_2$B$_2$C, and MgB$_2$", *Phys. Rev. B* **64** 180506(R) (2001)

[35] M. Bhatia, M.D. Sumption, E.W. Collings, and S. Dregia, "Increases in the Irreversibility Field and the Upper Critical Field of Bulk MgB$_2$ by ZrB$_2$ Addition", *Appl. Phys. Lett.* **87** 042505 (3pp) (2005)

[36] M. Bhatia, M.D. Sumption, S. Bohnenstiehl, S.A. Dregia, E.W. Collings, M. Tomsic, and M. Rindfleisch, "Superconducting properties of SiC doped MgB$_2$ formed below and above Mg's melting point", *IEEE Trans. Appl. Supercond.* **17** 2750-2753 (2007)

[37] V. Braccini, L.D. Cooley, S. Patnaik, D.C. Larbalestier, P. Manfrinetti, A. Palenzona, and A.S. Siri, "Significant enhancement of irreversibility field in clean-limit MgB$_2$", *Appl. Phys. Lett.* **81** 4577-4579 (2002)

[38] M. Mudgel, V.P.S. Awana, H. Kishan, and G.L. Bhalla, "Significant improvement of flux pinning and irreversibility field in *nano*-carbon-doped MgB$_2$ superconductor", *Solid State Comm.* **146** 330-334 (2008)

[39] H. Yamada, H. Yamasaki, K. Develos-Bagarinao, Y. Nakagawa, et al., "Flux Pinning Properties of c-Axis Correlated Pinning Centres in PLD-YBCO Films", *Supercond. Sci. Technol.* **17** S25-S29 (2004)

[40] T. Fujiyoshi, K. Toko, T. Matsushita, and K. Yamafuji, "Temperature Dependence of Flux Creep in High-$T_c$ Superconductors", *Japan J. Appl. Phys.* **28 L** 1906 – **L** 1908 (1989)

[41] K. Yamafuji, T. Fujiyoshi, K. Toko, and T. Matsushita, "A Theory of Thermally Activated Flux Creep in Nonideal Type II Superconductors", *Physica C* **159** 743-799 (1989)
27

## List of Tables

Table 1.

| B (T) | $U_{eff}$, meV (magn decay) | $U_{eff}$, meV ($n$-value) |
|---|---|---|
| 1 | 178 | - |
| 2 | 178 | - |
| 3 | 71 | - |
| 4 | 42 | - |
| 5 | - | 12.2 |
| 6 | - | 9.1 |
| 8 | - | 6.4 |
| 9 | - | 4.7 |
| 10 | - | 3.8 |

**List of Figures**

Fig. 1. (a) SEM (1000X) of a longitudinal section of the superconducting core of the present in-situ-processed PIT strand, (b) an SEM (750X) of its transverse section, (c) a high magnification transverse fracture SEM ($10^5$X) showing the random polycrystalline grain structure.

Fig. 2. Magnetic $J_c$ vs $B$ at temperatures between 4.2 K and 40 K for perpendicular and parallel applied fields.

Fig. 3. "Rooftop" critical state profiles and schematic magnetic $J_c$s for parallel and perpendicular orientations of the applied field.

Fig. 4. Magnetic and transport critical current densities in perpendicular applied fields for various temperatures.

Fig. 5. Magnetic vs transport $J_c$ in perpendicular applied fields evaluated for five electric-field criteria.

Fig. 6. "Rooftop" critical state profiles and schematic field-perpendicular magnetic and transport $J_c$s.

Fig. 7. *Main figure:* temperature dependence of resistivity in a 5 T field applied perpendicular and parallel to the strand axis using a DC current of 5 mA. *Inset: R versus T* in a perpendicular 5 T applied field using AC currents of 14 mA and 140 mA.

Fig. 8. *B-T* space, with *B* applied perpendicular to the strand axis, showing $B_{c2}$ and $B_{irr}$ as determined resistively and magnetically.

Fig. 9. *M-T* curves at various levels of field applied perpendicular to the strand axis.

Fig. 10. (a) Log-Log *E/J* curves based on Figure 5 data, (b) Magnetization decay in perpendicular applied fields at 20 K.

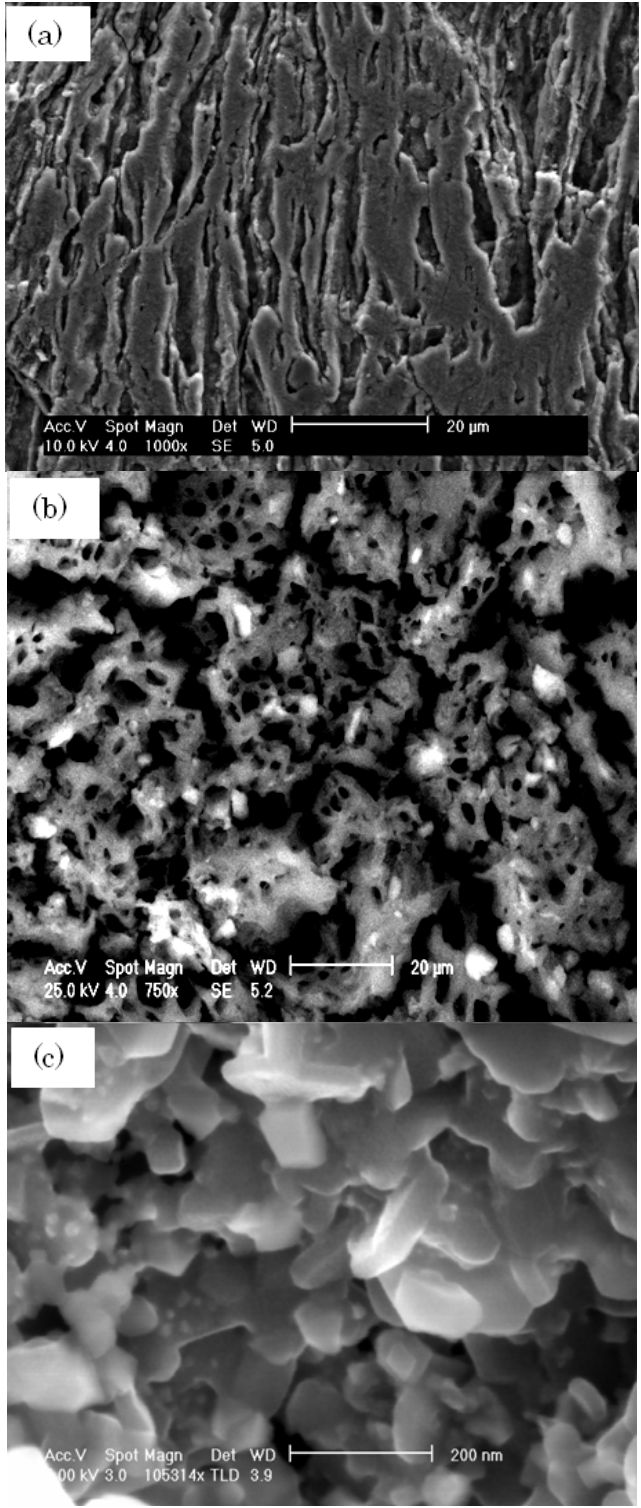

Fig. 1.

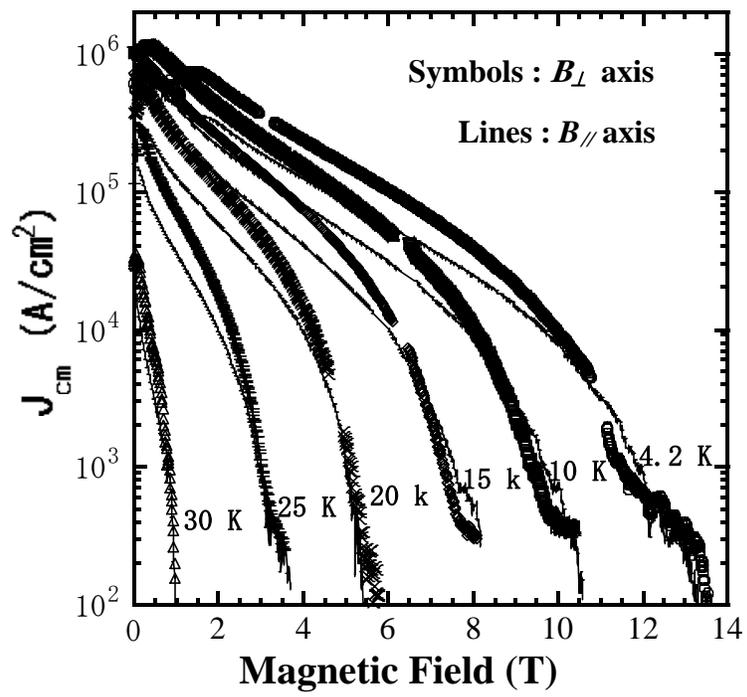

Fig. 2.

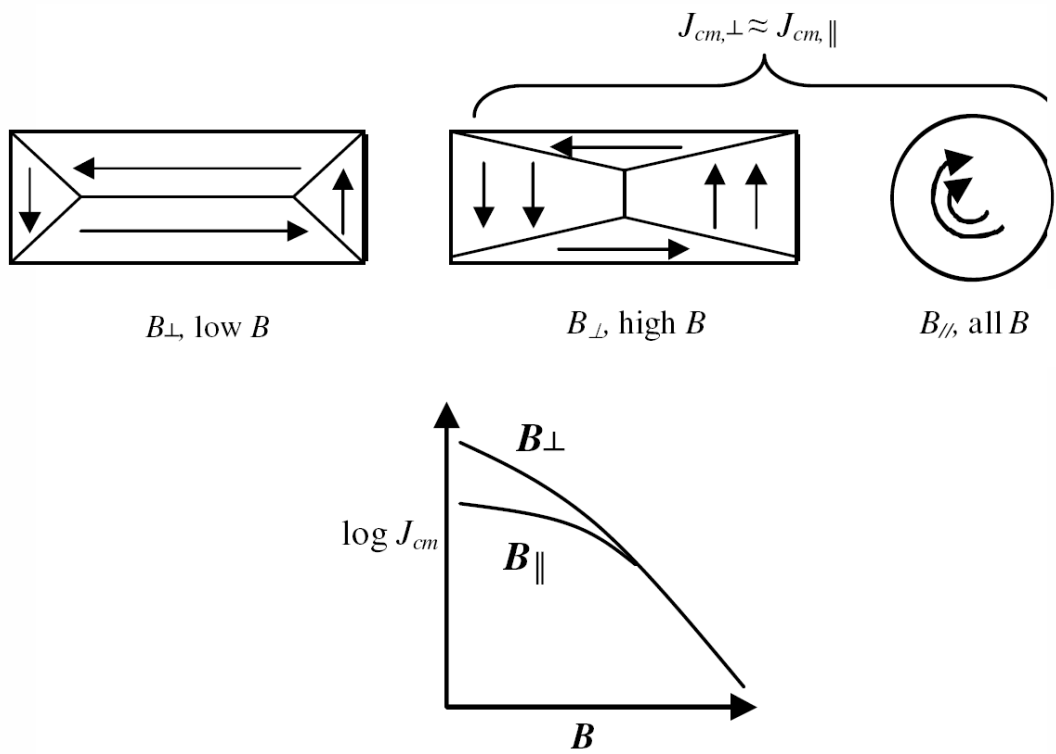

Fig. 3.

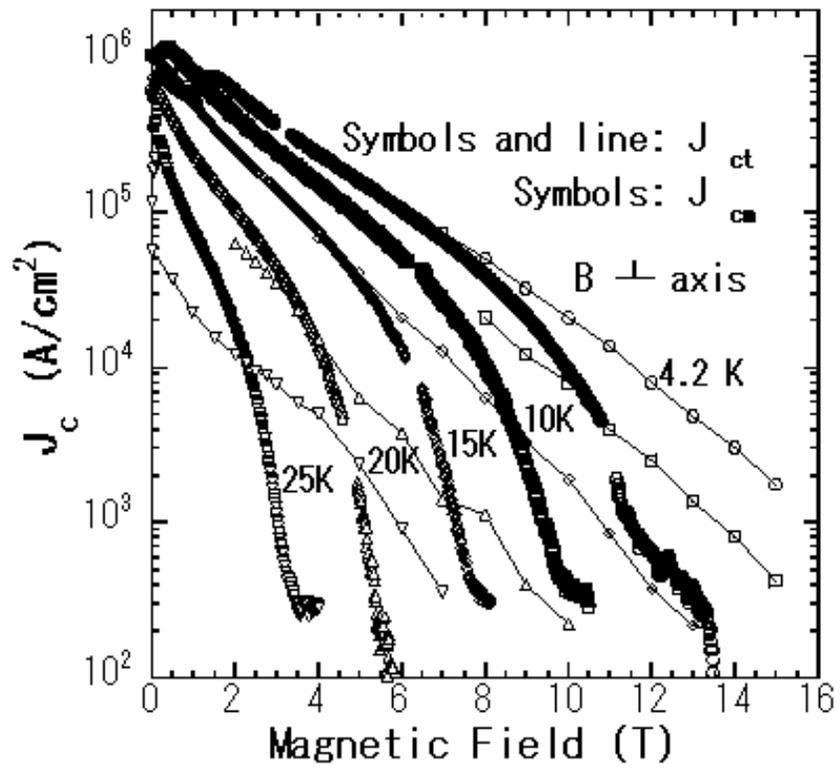

Fig. 4

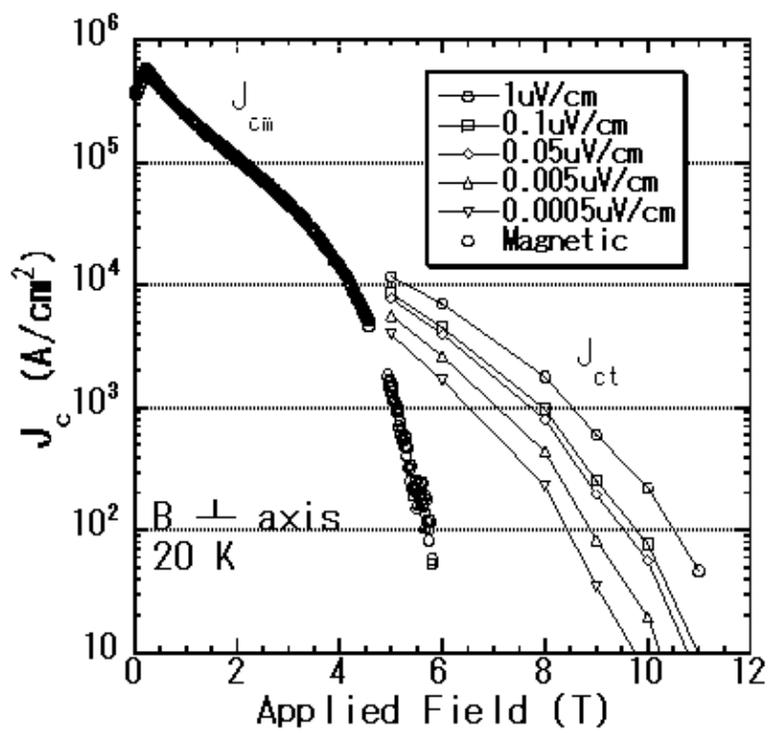

Fig. 5.

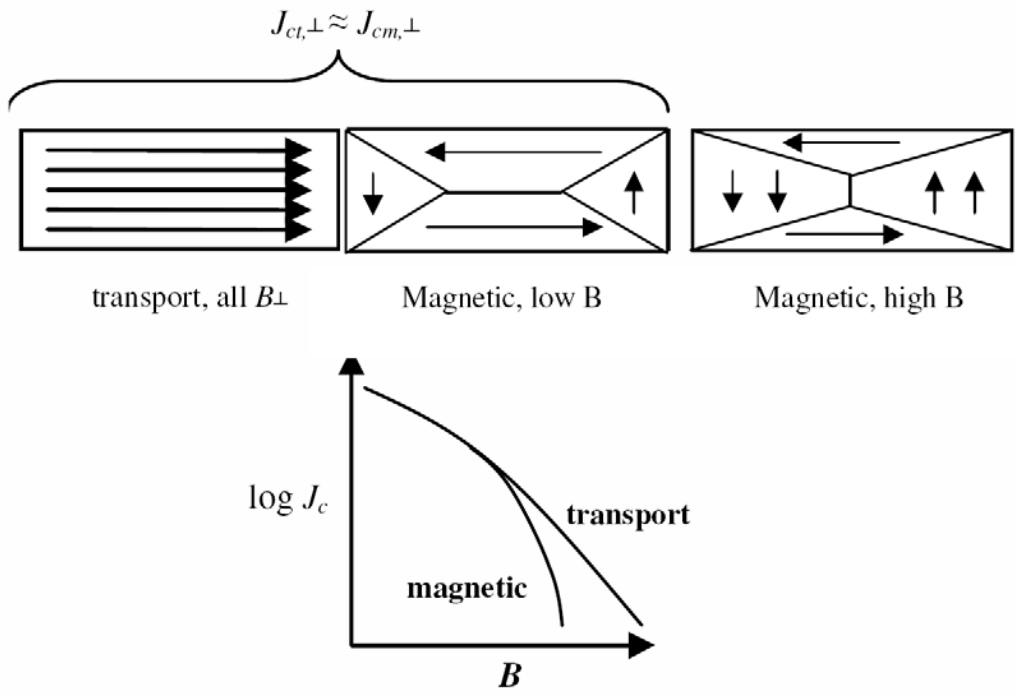

Fig. 6.

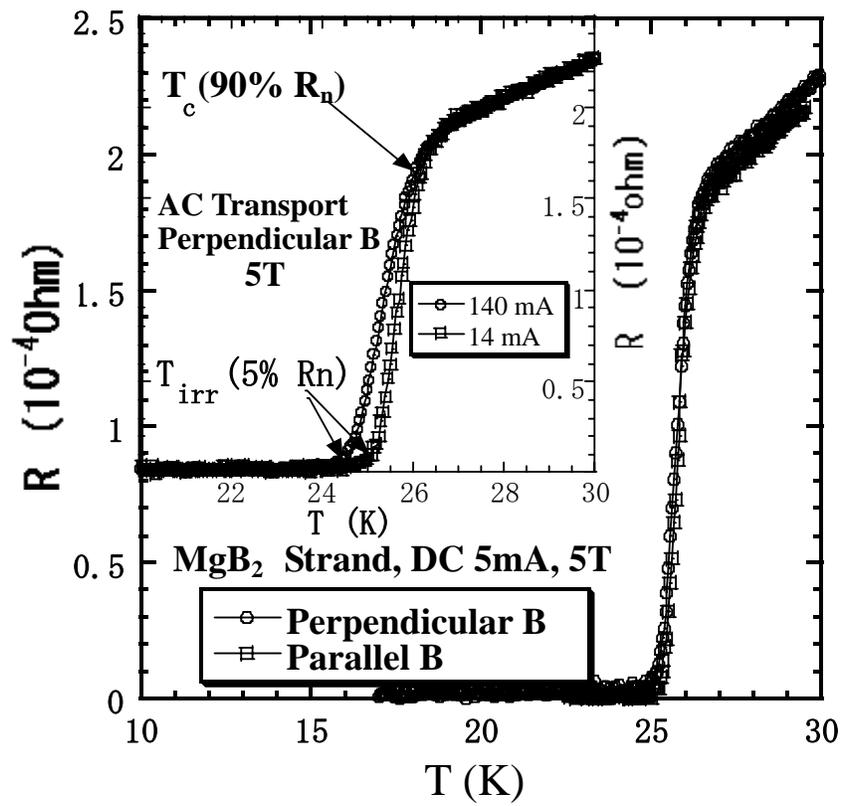

Fig. 7.

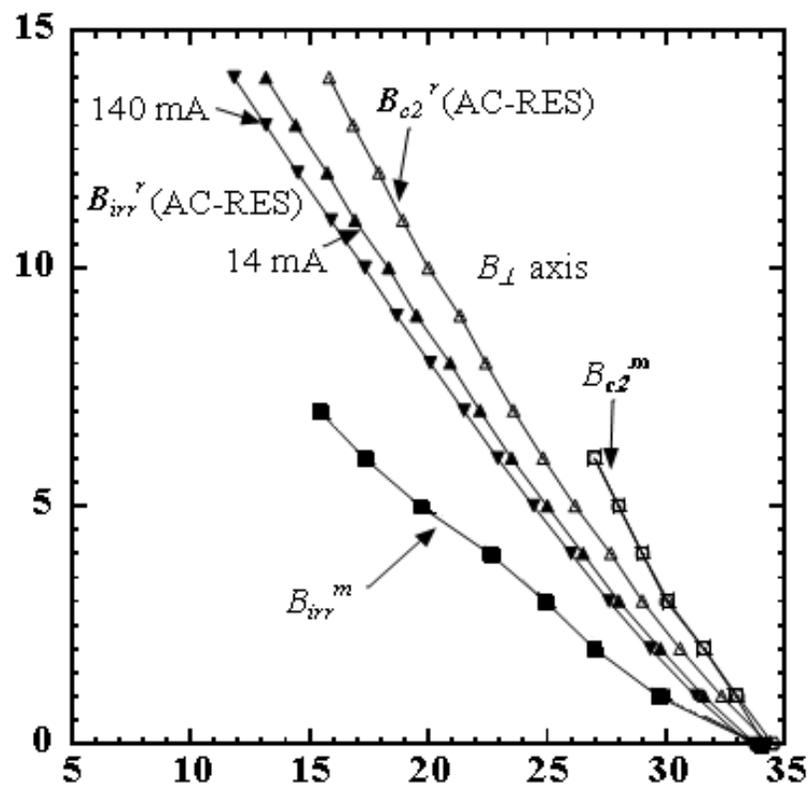

**Fig. 8.**

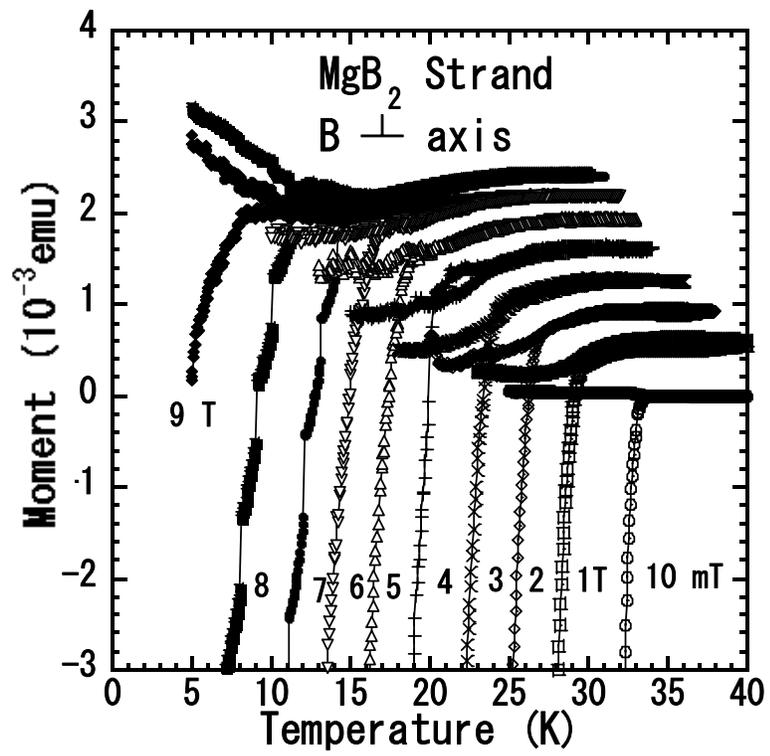

Fig. 9.

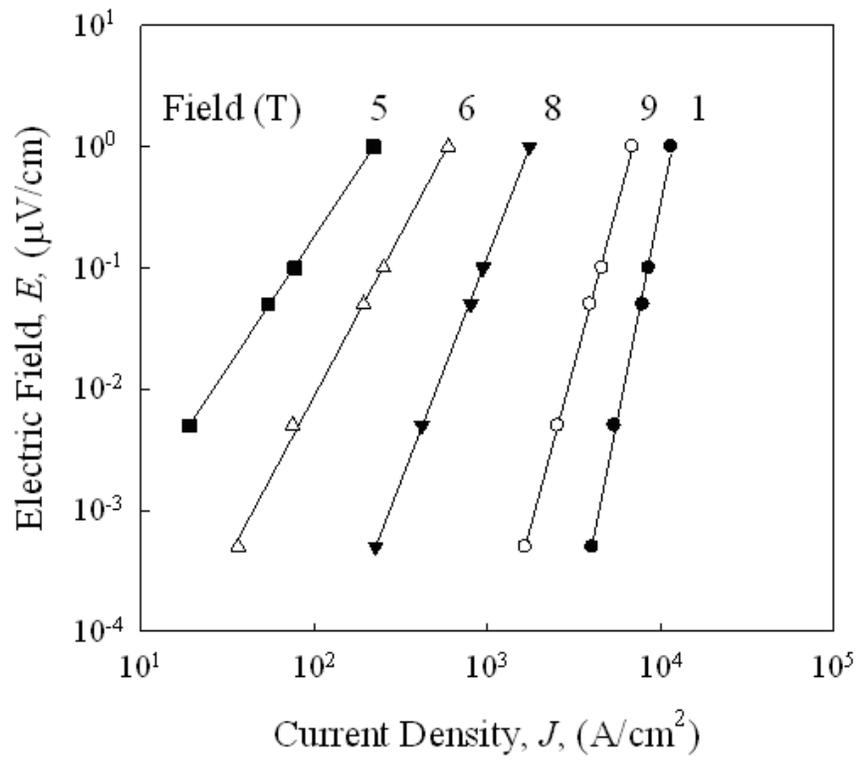

Fig. 10 (a).

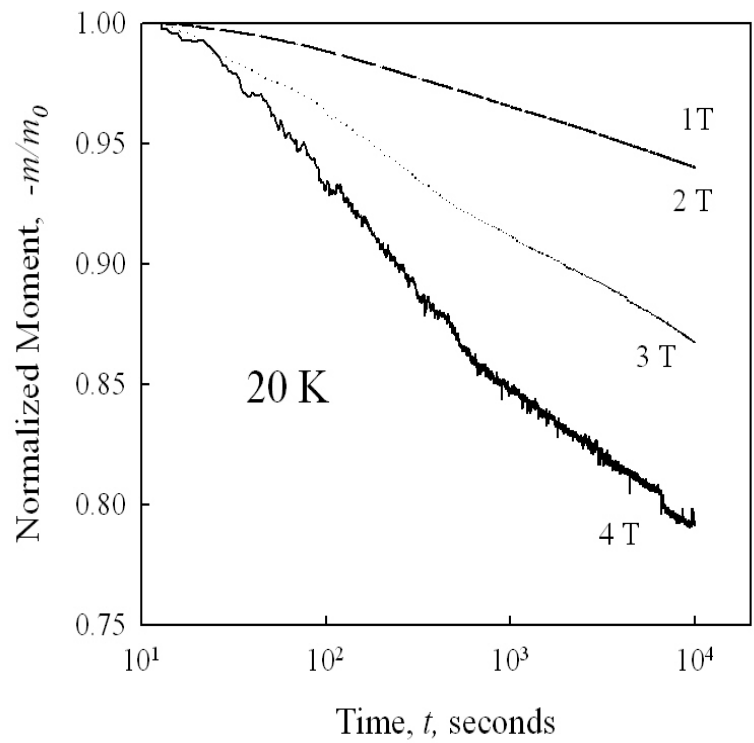

Fig. 10 (b).